\documentclass[aps,prl,amsfonts,amssymb,notitlepage,tightenlines,twocolumn,showpacs]{revtex4-1}
\usepackage[intlimits]{amsmath}
\usepackage{graphicx}

\usepackage{braket}

\newcommand{\rr}{\mathbf{r}}
\newcommand{\kk}{\mathbf{k}}
\newcommand{\HH}{\mathcal{H}}

\newcommand{\bosi}{\boldsymbol{\sigma}}

\begin{document}

\title{Optical signatures of states bound to vacancy defects in monolayer MoS$_2$}

\author{Mikhail Erementchouk $^{(1)}$}
\author{M. A. Khan $^{(1,2)}$}
\author{Michael N. Leuenberger $^{(1)}$}
\email[e-mail: ]{michael.leuenberger@ucf.edu}
\affiliation{$^{(1)}$ NanoScience Technology Center and Department of Physics, University
of Central Florida, Orlando, FL 32826, USA. \\
$^{(2)}$ Federal Urdu University of Arts, Science and Technology, Islamabad, Pakistan.}

\begin{abstract} 
We show that pristine MoS$_2$ single layer (SL) exhibits two bandgaps $E_{g\parallel}=1.9$ eV and $E_{g\perp}=3.2$ eV for the optical in-plane and out-of-plane susceptibilities $\chi_\parallel$ and $\chi_\perp$, respectively.
In particular, we show that odd states bound to vacancy defects (VDs) lead to resonances in $\chi_\perp$ inside $E_{g\perp}$ in MoS$_2$ SL with VDs.
We use density functional theory, the tight-binding model, and the Dirac equation to study MoS$_2$ SL with three types of VDs: (i) Mo-vacancy, (ii) S$_2$-vacancy, and (iii) 3$\times$MoS$_2$ quantum antidot. The resulting optical spectra identify and characterize the VDs.
\end{abstract}

\maketitle

\paragraph{Introduction.}
Monolayer transition metal dichalcogenides (TMDCs) (MX$_2$; M= transition metal such as Mo, W and  X=S, Se, Te) have attracted a lot of attention due to their intriguing electronic properties. Monolayer TMDCs are semiconductors with direct bandgap $E_{g\parallel}$ in the visible range, which makes them suitable for optoelectronic, spintronic, valleytronic, and photodetector devices \cite{Splendiani_2010,Mak_2010,Zhu_2011,xiao_coupled_2012,Lopez-Sanchez_2013,liu_three_band_2013,rostami_effective_2013}. In order to increase the performance of such devices based on TMDC single layer (SL), it is crucial to characterize the defects present in TMDC SLs.

\begin{figure*}[bt]
	\begin{center}
		\includegraphics[width=7in]{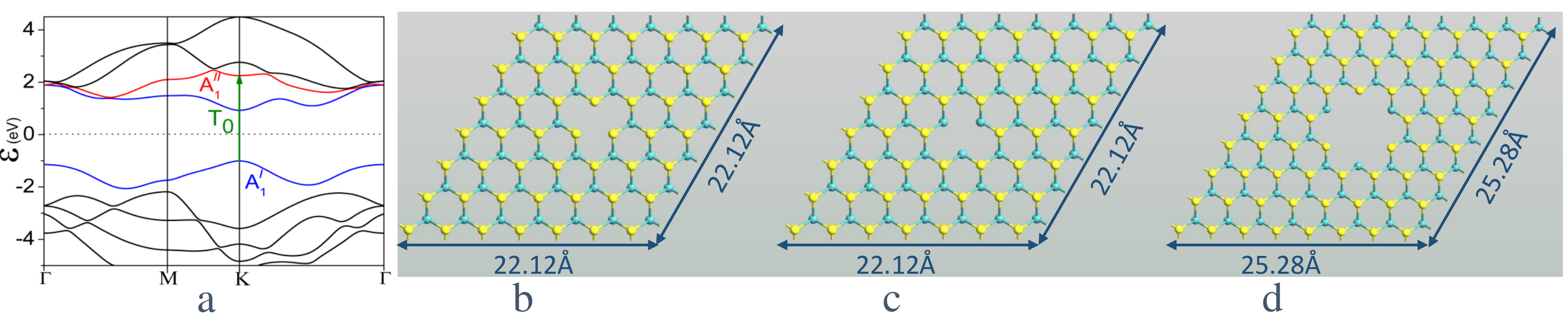}
	\end{center}
	\caption{(a) Bandstructure of MoS${}_2$ SL, showing the out-of-plane bandgap $E_{g\perp}=3.2$ eV determined by the transition $T_0$. The Fermi level is set at $\epsilon_F=0$ eV. (b) Mo-vacancy in 7x7 supercell, (c) S$_2$-vacancy consisting of a pair of S atoms removed in 7x7 supercell, (d)  hexagonal 3$\times$MoS$_2$ QAD in 8x8 supercell.}
	\label{fig:structure}
\end{figure*}

Here we show that the bandgap $E_{g\perp}=3.2$ eV for the optical out-of-plane susceptibility $\chi_\perp$ in pristine MoS$_2$ SL provides a large energy window to characterize vacancy defects (VDs). Pristine MoS$_2$ SL is invariant with respect to $\sigma_h$ reflection about the $z=0$ (Mo) plane, where the $z$ axis is oriented perpendicular to the Mo plane. Therefore, electron states break down into two classes: even and odd, or symmetric and anti-symmetric with respect to $\sigma_h$. We show below that this leads to the nontrivial consequence that $\chi_z=\chi_\perp$ has a bandgap of $E_{g\perp}=3.2$ eV, which is substantially larger than the bandgap $E_{g\parallel}=1.9$ eV for  the in-plane component of the optical susceptibility $\chi_x=\chi_y=\chi_\parallel$. As we show, due to the optical selection rules for the even and odd states, there are no $\pi$ transitions, driven by $z$-polarized photons, below $3.2$ eV. Hence, $\chi_\perp$ for pristine MoS$_2$ SL must vanish for energies below $3.2$ eV. 

Several studies on VDs in 2D materials have emerged. The minibands resulting from quantum antidot (QAD) superlattices can be used to tune the bandgaps of graphene \cite{Pedersen_2008} and MoS$_2$ SL \cite{Huang_2013,shao_theoretical_2014}.
In another study, we have shown that substitutional defects in the form of MoO$_3$ not only lead to strong suppression of the conductivity \cite{Islam_2014} but also to photoluminescence quenching \cite{Kang_2014}. Recently, VDs in MoS${}_2$ SL have been characterized theoretically in terms of magnetic properties \cite{Zhou_2013}. A recent experimental study used scanning transmission electron spectroscopy to characterize several types of defects in MoS$_2$ SL, including Mo, S, and S$_2$ VDs \cite{Hong_2015}. MoS$_2$ SL with S-vacancies might catalyze alcohol synthesis from syngas \cite{Le_2014}.

Here we show that VDs yield strong resonances in $\chi_\perp$, which provides the opportunity to optically characterize VDs in MoS$_2$ SL with VDs (denoted by MoS$_2$ SLVD). 
We consider the optical signatures of states bound to three types of VDs in MoS$_2$ SLVD: (i) Mo-vacancy, (ii) S$_2$-vacancy, and (iii) a hexagonal 3$\times$MoS$_2$ QAD (see Fig. \ref{fig:structure}).

\begin{figure}[b]
	\begin{center}
		\includegraphics[width=3.5in]{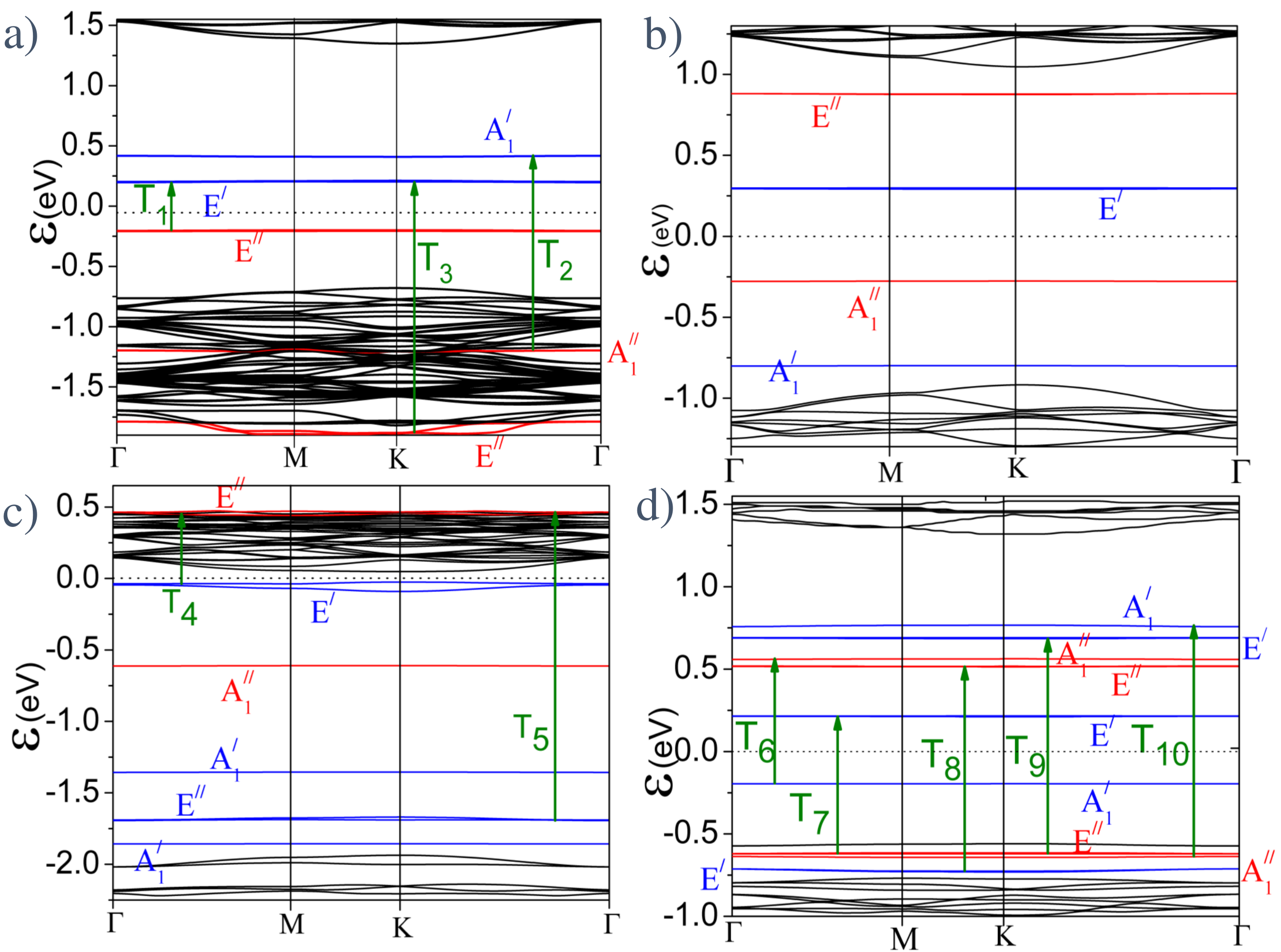}
	\end{center}
	\caption{Bandstructure of different kinds of VDs. The Fermi level is set at $\epsilon_F=0$ eV. Red (blue) lines show odd (even) states. Arrows indicate transitions corresponding to resonances in $\chi_\perp(\omega)$ (see Eq.~\eqref{eq:chi_KG} and below) shown in Fig.~\ref{fig:optical}. (a) Mo-vacancy. (b) S$_2$-vacancy. Here singlets and doublets are in different half-planes with respect to the Fermi level. Therefore there are no $\pi$ transitions. (c) S$_2$-vacancy with charge added to MoS$_2$ SLVD, raising the Fermi level such that $\pi$ transitions become allowed. (d) 3$\times$MoS${}_2$ QAD.}
	\label{fig:band_structure}
\end{figure}

\paragraph{Bandstructure.}
First we start with the numerical bandstructure calculation of MoS${}_2$ SLVD using standard Density Functional Theory (DFT) with meta-GGA functionals \cite{Tao_2003}, providing accurate estimates of bandgaps without the need to perform computationally intensive DFT calculations using the GW approximation \cite{GW_1,GW_2}. The calculations are implemented within Atomistix Toolkit 2014.2 \cite{QW_1}. The resulting bandstructures are shown in Fig.~\ref{fig:band_structure}. The periodic structure of the superlattice allows one to characterize the electron states by the bandstructure $\epsilon_n(\mathbf{k})$, where $\mathbf{k}$ is the vector in the first Brillouin zone of the superlattice and $n$ enumerates different bands. We consider supercells with dimensions $7\times7\times1$ (Fig.~\ref{fig:band_structure} b, c) and $8\times8\times1$ (Fig.~\ref{fig:band_structure} d) having $147$ and $192$ number of atoms, respectively. For Brillouin zone integration we consider $k$ sampling of $7\times7\times1$. The cut off energy is set to $300$ eV and the structure is optimized by using a force convergence of $0.01$ eV/\AA.

\paragraph{Tight-binding model (TBM) and symmetries.}
Within the TBM approximation the electron wavefunction can be presented as
%\begin{equation}\label{eq:wave_TBM}
 $\Ket{\psi} = \sum_{j,\mu \in O_j} \psi_\mu^{(j)} \varphi_\mu^{(j)}(\rr - \mathbf{R}^{(j)})$,
%\end{equation}
where $j$ enumerates atoms in the layer and the summation over $\mu$ runs over respective atomic orbitals, whose set for the $j$-th atom is denoted $O_j$. Choosing $\mathbf{e}_{x,y}$ in the plane of the layer and $\mathbf{e}_z$ perpendicularly, 
for Mo the real orbitals of main importance are the $d$-orbitals $d_{x^2 - y^2}$, $d_{xy}$ and so on, while for S atoms these are $p$-orbitals $p_i^{(t,b)}$ with $i = x, y, z$ and $t$ and $b$ denoting the top and bottom layers, respectively. The classification of the electron states simplifies when the symmetry with respect to $\sigma_h: z \mapsto -z$ is taken into account. The electron states transform according to $A$ and $A'$, the irreducible representations of $Z_2 = \{e, \sigma_h \}$. The respective even and odd orbitals are locally spanned by the bases \cite{cappelluti_tight-binding_2013}: $\{d_{x^2 - y^2}, d_{xy}, d_{z^2},(p_{x,y}^{(t)} + p_{x,y}^{(b)})/\sqrt{2}, (p_{z}^{(t)} - p_{z}^{(b)})/\sqrt{2} \}$  and $\{d_{xz}, d_{yz}, (p_{x,y}^{(t)} - p_{x,y}^{(b)})/\sqrt{2}, (p_{z}^{(t)} + p_{z}^{(b)})/\sqrt{2} \}$.

The full group of the point symmetries of MoS${}_2$ SLVD with our considered VDs is $D_{3h} = C_{3v}\otimes Z_2$. Thus the states bound to the VDs, even and odd with respect to $\sigma_h$, must transform according to $A_{1,2}$ and $E$, the irreducible representations of $C_{3v}$. Respectively, the bound states must appear as singlets and doublets.
It should be noted that such classification holds if the overlap between states bound to different VDs is absent.
% as is illustrated by Figs.~\ref{fig:band_structure}a and \ref{fig:band_structure}b.

The simplest model describing the arrangement of the bound states is the TBM considering only the atoms on the edge of the VD, which is inferred by the small localization radius of the bound states. For the case of the hexagonal $3\times$MoS$_2$ QAD one finds
\begin{equation}\label{eq:energies_plaquette}
\epsilon = \bar{\epsilon} \pm \sqrt{\delta \epsilon^2 + 4 |t|^2 \cos^2(\xi)},
\end{equation}
where $\bar{\epsilon} = \left(\epsilon_{Mo} + \epsilon_{S} \right)/2 $, $\delta{\epsilon} = \left(\epsilon_{Mo} - \epsilon_{S} \right)/2 $ and $\xi = 0$ for singlet states (invariant with respect to $C_3$ rotations) and $\xi = 2\pi/3$ for doublets (states aquiring the phase factor $\exp(\pm i 2\pi/3) $). Here $\epsilon_{Mo}$, $\epsilon_S$ are the phenomenological parameters describing the energy of the electron on Mo and S atoms, respectively, and $t$ is the hopping parameter. Equation~\eqref{eq:energies_plaquette} correctly reproduces the sequence $\{A_1', E', E', A_1'\}$ for even and $\{A_1'', E'', E'', A_1''\}$ for odd states, i.e. $\{$singlet, doublet, doublet, singlet$\}$, while traversing the gap $E_{g\parallel}$ from the bottom of the conduction band down over the bound states (see Fig.~\ref{fig:band_structure}d).

\begin{figure}[tb]
	\begin{center}
		\includegraphics[width=3.5in]{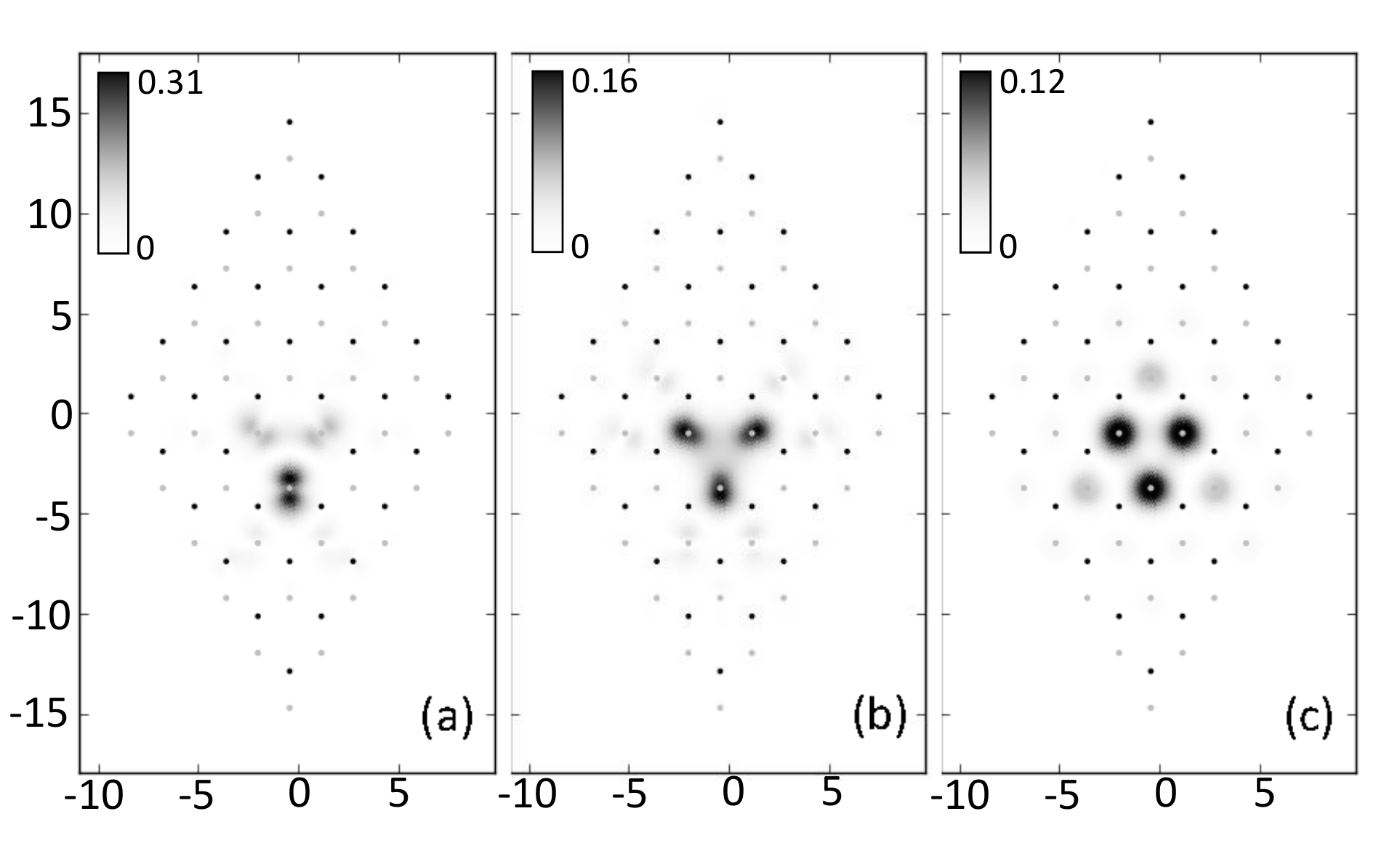}
	\end{center}
	\caption{Examples of the electron probability distributions in states bound to Mo-vacancy. Single super-cell is shown, black and gray dots indicate positions of Mo and S atoms, respectively. The distances are measured in \AA. (a) Odd doublet at energy $\approx 0.5$ eV above the top of the valence band, (b) odd singlet inside the valence band at energy $\approx 0.5$ eV below the top of the valence band, (c) deep defect state at energy $\approx -12$ eV below the Fermi level formed by Sulfur's $s$ orbitals.}
	\label{fig:new_states}
\end{figure}

It may appear that this model contradicts the numerical results for the Mo-vacancy, where the numerical calculations show only $5$ bound states (see Fig.~\ref{fig:band_structure}a). However, the TBM model suggests that in addition to the bound states appearing inside the gap $E_{g\parallel}$  of MoS${}_2$ SL there must be states, in this particular case a singlet state, hidden inside the bands. Indeed, there is such a state inside the valence band at energy $\epsilon \approx -0.5$ eV below the top of the valence band (see Fig.~\ref{fig:new_states}b). 

Moreover, the parameters $\epsilon_{Mo} $ and $\epsilon_{S} $ are determined by the microscopic Hamiltonian, e.g. $\epsilon_{S} = \Braket{\phi^{(S)}_\mu | \HH | \phi^{(S)}_\mu} $. Thus, we can expect that there is a variety of states bound to VDs besides the ones inside the gaps $E_{g\parallel}$, $E_{g\perp}$. An example of such states is provided by the case when the bound state is made of Sulfur's $s$-orbitals (see Fig.~\ref{fig:new_states}c) at energy $\approx -12$ eV below the Fermi level.

\paragraph{Dirac model.}
The TBM considered above relates the structure of the spectrum of the bound states to the symmetry of the VD. Due to the fact that its parameters should be fitted to the energies of electron states obtained by other means,  however, it cannot explain neither the smallness of the localization radius of the bound states nor their energies. In particular, it cannot explain why the odd states may form bound states inside the gap $E_{g\parallel}$. These features, however, can be understood with the help of an analysis of circularly symmetric QADs based on the Dirac equation which emerges as the two-band model within the $k\cdot p$-approximation near the $K$-point of the Brillouin zone of MoS$_2$ SL. 

Considering two bands with the energy separation $2\Delta$ between them, the equation describing the spatial distribution of the pseudo-spin has the form $\HH_\tau \Phi_\tau = \epsilon \Phi_\tau$, where $\tau = \pm 1$ enumerates the valleys, the energy reference level is chosen to be positioned at the center between the bands, and
%\begin{equation}\label{eq:Ham_tau}
$\HH_\tau = \tau \sigma_z \Delta + v \bosi \cdot \mathbf{p}$.
%\end{equation}
Assuming that the QAD has circular shape, we rewrite this equation in the polar coordinates and with respect to $\widetilde{\Phi}_\tau = \exp(i \sigma_z \phi/2)\Phi_\tau \sqrt{r} $ we obtain $\widetilde{\HH}_\tau \widetilde{\Phi}_\tau = \epsilon \widetilde{\Phi}_\tau$, where
\begin{equation}\label{eq:H_rot}
\widetilde{\HH}_\tau = \tau \sigma_z \Delta - i v \left[\sigma_x \frac{\partial}{\partial r}  + \sigma_y \frac{1}{r}\frac{\partial}{\partial \phi} \right].
\end{equation}
The solution is subject to the condition of vanishing radial component of the probability current at the boundary \cite{mccannSymmetry2004,akhmerovBoundary2008} $\Braket{\Phi(r_0) | \mathbf{n}_B \cdot \bosi | \Phi(r_0)} = 0$, where $r_0$ is the radius of the QAD and $\mathbf{n}_B$ is the unit vector perpendicular to the boundary. The straightforward implementation of such boundary condition is provided by the infinite mass model \cite{berryNeutrino1987}, where the QAD is represented by a region with renormalized width of the gap $\Delta \to \Delta(1+d(r))$, with $d(r) = 0$ for $r> r_0$ and $d(r) \to \infty$ when $r< r_0$. In our case we identify $\Delta=E_{g\parallel}$. Within this model the boundary condition is satisfied if $\Ket{\Phi(r_0)} \propto \Ket{\tau y} $, i.e. the pseudo-spin $\Ket{\Phi(r_0)}$ is tangent to the boundary of the QAD. Next, observing that $\sigma_z \Ket{\tau y} = \Ket{-\tau y} $ and $\sigma_x \Ket{\tau y} = i \tau \Ket{-\tau y} $, one can see that $\widetilde{\HH}_\tau $ has an angularly independent solution $\Ket{\Phi(r_0)} \propto \exp(- r/r_c) \Ket{\tau y} $ corresponding to $\epsilon = 0$, which exponentially decays for $r > r_0$ with the localization radius $r_c = \Delta/v$.

Thus, independently of its radius the QAD may support a bound state with very short localization length and with the energy in the middle between the energies of the coupled bands. Comparing this finding to the distribution of energies of even and odd bands \cite{cappelluti_tight-binding_2013} the conclusion can be drawn that, indeed, both even and odd bands may support bound states with the energy near the energy of the Fermi level, i.e. inside the gap $E_{g\parallel}$ of MoS$_2$ SL.

\paragraph{Optical spectrum.}
In view of nontriviality of the appearance of odd bond states inside the gap $E_{g\parallel}$ it is important to note that the presence of the bound states of different parities manifests itself in the optical spectrum of MoS${}_2$ SLVD. Therefore, they are available for a direct experimental observation.

When VDs form a superlattice the problem of the optical response can be approached along the same line as for single layered systems \cite{rose_spin_2013}. 
Let $\kk$ be a point in the first Brillouin zone of the superlattice. At this point the electron wave function satisfies
\begin{equation}\label{eq:nk_eq}
 \epsilon_n(\kk) \Ket{\psi_n(\kk)} = \HH(\kk) \Ket{\psi_n(\kk)},
\end{equation}
where $n$ enumerates the superlattice bands. 
Implementing the $k\cdot p$-approximation of Eq.~\eqref{eq:nk_eq} in the usual way and using the Peierls substitution we obtain the Hamiltonian of interaction with the elecromagnetic field 
%\begin{equation}\label{eq:H_EM}
$\HH_{EM} = \frac{e}{m} \mathbf{A} \cdot \mathbf{p}$.
%\end{equation}
Treating $\HH_{EM}$ as a perturbation within the linear response theory we find the Kubo-Greenwood optical susceptibility (see e.g. Ref.~\cite{harrison_solid_1970})
\begin{equation}\label{eq:chi_KG}
\begin{split}
\widehat{\chi}(\omega) = \frac{\pi e^2 \hbar^4}{m^2 \epsilon_0 \omega^2} & \sum_{n,n'} \int d\kk  \,
	\mathbf{P}_{n,n'}(\kk) \otimes \mathbf{P}_{n',n}(\kk) \\ 
 &
 \times \frac{f[\epsilon_n(\kk)] - f[\epsilon_{n'}(\kk)]}{\epsilon_n(\kk) - \epsilon_{n'}(\kk) -\omega - i \gamma},
\end{split}
\end{equation}
where $f(\epsilon)$ is the Fermi distribution, $\otimes$ denotes the tensor product and 
$\mathbf{P}_{n,n'}(\kk) = \Bra{\psi_n(\kk)} \mathbf{p} \Ket{\psi_{n'}(\kk)}$.

The appearance of the states inside the gap $E_{g\parallel}$ of MoS${}_2$ SL leads to resonances at frequencies of corresponding transitions. Several transitions, however, are prohibited due to symmetry, i.e. when $\mathbf{P}_{n,n'}(\kk)$ does not transform according to the symmetric representation of the symmetry group of the superlattice. 
$\mathbf{P}_{n,n'}(\kk)$ transforms according to $I(C_{3v})^2 \otimes I(Z_2)^2 \otimes I(\mathbf{P}) $, where $I(G)$ and $I(\mathbf{P})$ denote irreducible representations of group $G$ and the momentum operator $\mathbf{P}$, respectively, and powers are shorthand notations for direct products. One needs to consider separately the in-plane and out-of-plane components of $\mathbf{P}_{n,n'} $ because they transform according to different irreducible representations of $D_{3h}$, namely, $E' = E \otimes A'$ and $A_2'' = A_1 \otimes A''$, respectively. Taking into account the multiplication rules for $C_{3v}$: $A_{1,2}^2 = A_1$, $A_{1,2} \otimes E = E$ and $E^2 = A_1 \oplus A_2 \oplus E$, we find that the out-of-plane component of $\mathbf{P}_{n,n'}(\kk)$, which gives rise to $\pi$ transitions, is nonzero only between odd and even states of the same multiplicity (either between singlets or between doublets), while the in-plane components, which lead to $\sigma$ transitions, are nonzero for all states of the same parity. Thus $\widehat{\chi}$ is diagonal in the basis spanned by $\mathbf{e}_{x, y, z}$ and is isotropic in the plane of the layer and, thus, is characterized fully by two eigenvalues $\chi_\parallel$ and $\chi_\perp$.

\begin{figure}[tb]
	\begin{center}
		\includegraphics[width=3.5in]{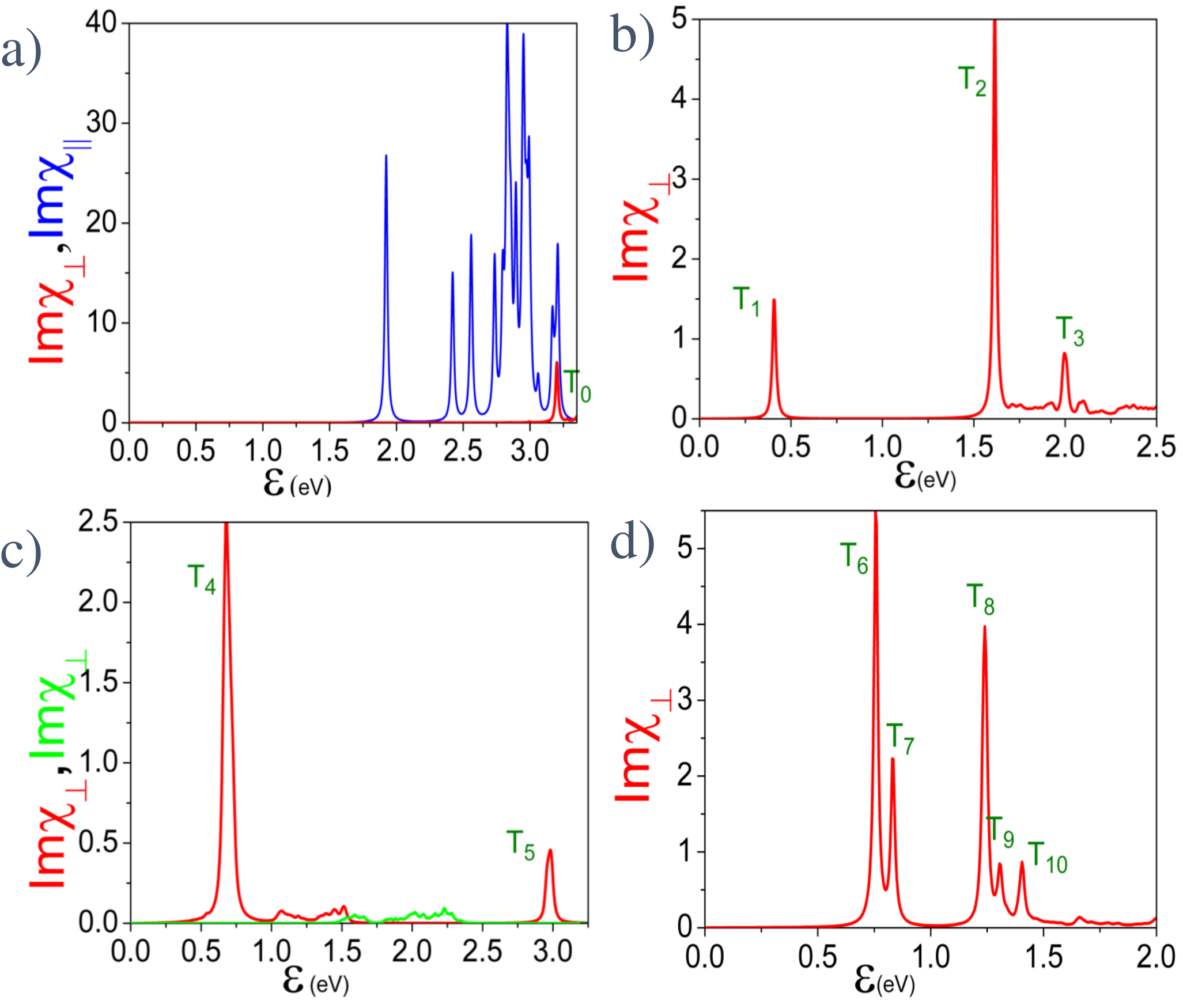}
	\end{center}
	\caption{(a) Resonances of $\mathrm{Im}\chi_\parallel(\omega)$ (blue) and $\mathrm{Im}\chi_\perp(\omega)$ (red) in pristine MoS$_2$ SL. (b) MoS${}_2$ SLVD with Mo-vacancy: $T_3$ resonance in $\chi_\perp(\omega)$ is due to the odd singlet state inside the valence band. (c) MoS${}_2$ SLVD with S$_2$-vacancy: If uncharged (green line), $\pi$ transitions are suppressed; if charged (red line), $\pi$ transitions are allowed. (d) MoS$_2$ SLVD with $3\times$MoS$_2$.}
	\label{fig:optical}
\end{figure}

The numerical results for the optical spectrum are shown in Fig.~\ref{fig:optical}.
The difference between $\chi_\perp(\omega)$ for pristine MoS${}_2$ SL and for MoS${}_2$ SLVD is drastic. For pristine MoS${}_2$ SL the lowest energy transition $T_0$ yielding nonzero $\chi_\perp$ (see Fig.~\ref{fig:optical}a) corresponds to the transition between the top of the valence band to the $\mathrm{CB}+1$ band with energy $3.2$ eV (see Fig.~\ref{fig:structure}a). In turn, for MoS${}_2$ SLVD the lowest energy resonance is due to the transition between bound states of the same degeneracy with the energy difference smaller than $1$ eV (see Fig.~\ref{fig:optical}b, c, and d). This result is in stark contrast to true 2D systems where $\pi$ transitions are absent, and the effect of non-zero small thickness may be expected to be observed at energies at least significantly higher than those characteristic to $\sigma$ transitions. In addition, the selection rules governing $\pi$ transitions present a great opportunity for experimental characterization of states bound to VDs. 

The qualitative picture based on the symmetry properties establishes the connection between the main features of spectrum of the bound states and the optical response. For example, for the Mo-vacancy the $T_3$ resonance in $\chi_\perp(\omega)$ involves a bound state hidden in the valence band. In the case of S$_2$-vacancy the symmetry analysis predicts that $\chi_\perp(\omega)$ is featureless at low energies due to the smallness of $f_f - f_i$, which is confirmed by the numerical calculations (see Fig.~\ref{fig:optical}b). 
The Fermi level $\epsilon_F$, however, can be shifted by means of a gate voltage to lie between equally degenerate states with different parities, which support transitions contributing to $\chi_\perp(\omega)$. Then $\chi_\perp(\omega)$ should demonstrate a low-energy resonance. In the numerical simulations we modified the position of the Fermi level by adding charge to the whole layer by means of a charge concentration of $1.4\times10^{21}$cm$^{-3}$, leading to a resonance in $\chi_\perp(\omega)$ (see Fig.~\ref{fig:optical}c). Fig.~\ref{fig:optical}d shows the resonances due to transitions between states bound to $3\times$MoS$_2$.

\begin{figure}[htb]
	\begin{center}
		\includegraphics[width=3.5in]{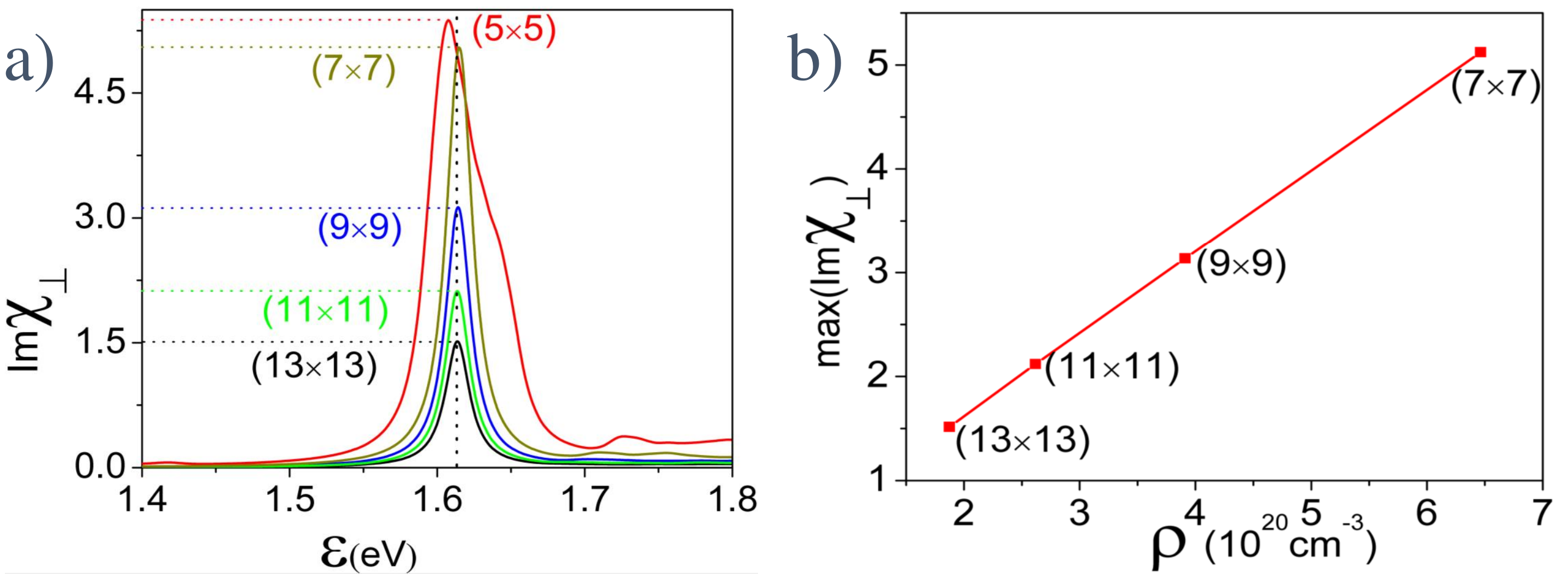}
	\end{center}
	\caption{(a) Magnitude of $\chi_\perp(\omega)$ as a function of VD concentration $\rho$ for Mo-vacancy around the peak $T_2$ (see Fig.~\ref{fig:optical}b) for the five supercells 5x5, 7x7, 9x9, 11x11, and 13x13. (b) Linear dependence of $\chi_\perp(\omega)$ at maximum of peak $T_2$.}
	\label{fig:magnitude}
\end{figure}

Considering the case when there is no overlap between bound states of neighboring VDs, we can use $\mathrm{Im}\chi_{\perp,T_i}=\rho(d_{T_i}^2/\epsilon_0\hbar)\gamma/[\gamma^2+(\epsilon_n - \epsilon_{n'}-\omega)^2]$ for the transition $T_i$ of a dilute gas of VDs \cite{Grynberg_2010}, where $\rho=N/V$ and $N$ are the concentration and number of VDs, respectively. $d_{T_i}= \Bra{\psi_{n_{T_i}}(\rr)} \rr \Ket{\psi_{n_{T_i}'}(\rr)}$ denotes the dipole moment of the transition $T_i$. This formula is in excellent agreement with the numerical calculations shown in Fig.~\ref{fig:magnitude} for supercell sizes from 7x7 up to 13x13. The peak for the 5x5 supercell does not follow this formula because the overlap between neighboring VDs is substantial, which leads to a peak shift and homogeneous peak broadening due to the formation of minibands. Additional inhomogeneous peak broadening is expected due to random distribution of VDs in the MoS$_2$ SL.

\paragraph{Conclusion.}
We show that in order to describe the electron states bound to VDs in MoS$_2$ SLVD, it is necessary to consider odd states, which lead to the appearance of resonances in the out-of-plane optical response $\chi_\perp(\omega)$. Our results pave the way to the optical characterization of VDs in TMDC SLVD, which is of utmost importance for the future realization of high-performance electronic and optoelectronic devices based on TMDC SLs.

\begin{acknowledgments}
\paragraph{Acknowledgments.}
We acknowledge support provided by NSF grant ECCS-1128597. We thank Saiful Khondaker and Laurene Tetard for useful comments.
\end{acknowledgments}

%\bibliography{Defect}


\end{document}